\documentclass[aps,prl,showpacs,twocolumn]{revtex4}
\usepackage{graphicx}
\begin{document}

\title{ Phason Disorder Effects in the Penrose Tiling Antiferromagnet }
\author{ Attila Szallas and Anuradha Jagannathan }
\affiliation{Laboratoire de Physique des Solides, CNRS-UMR 8502, Universit\'e
Paris-Sud, 91405 Orsay, France }

\date{\today}

\begin{abstract}
We discuss the ground state of a disordered two dimensional Heisenberg antiferromagnet. The starting structure is taken to be a perfectly deterministic quasiperiodic tiling, and the type of disorder we consider is geometric, involving frozen phason flips of a randomly selected subset of sites. We consider S=1/2 quantum spins placed on the vertices of the tiling, and interacting with the nearest neighbor spins with a uniform exchange interaction J. We calculate the energy spectrum, ground state energy and real space local magnetization values as a function of degree of disorder. We find that quantum fluctuations are enhanced by disorder. The real space staggered magnetization loses its symmetry properties and the average staggered magnetization decreases compared to the case of the perfect Penrose tiling. We explain our results in terms of a simple Heisenberg star cluster model that takes into account the changes of local environments due to phason flips. 
\end{abstract}

\maketitle

It is well-known that the electronic properties of random tilings are qualitatively different from those of the perfect quasiperiodic tiling \cite{sire}. To cite one example, the perfect quasiperiodic system tends to have strongly singular fluctuations in the density of states, while the random systems have fluctuations that are similar in some respects to those present in weakly disordered periodic systems. We consider a disordered magnetic model : the Heisenberg antiferromagnet defined in disordered approximants of the two dimensional Penrose tiling \cite{penref,duneaukatz}. We investigate the excitation spectrum, the ground state energy and staggered local magnetizations in these systems and compare the results with the perfect deterministic case (see \cite{penroselong, penro, ICQ10_Anu}). We create a frozen disorder in the perfect deterministic tiling by making randomly located phason flips. Phason flips are operations that reorganize the structure locally, in the vicinity of the flipped site. This type of disorder is strongly constrained, as one does not modify the basic building blocks of the structure, but only the way they are put together. Many structurally refined quasicrystals are assumed to be basically deterministic, with phonon and phason disorder at finite temperatures \cite{levine}. However, models for basically random structures are also proposed for a second category of quasicrystals, based on arguments related to maximising entropy of the system \cite{boissieu_icq9}. Many experimental studies, using neutron or X-ray diffraction have been carried out to detect phason disorder and compare with existing theoretical models. There are many studies of physical properties of random quasiperiodic structures and the influence of phason disorder. For example, it is interesting to consider the effect of disorder on transport properties, where some studies have shown that the electrical conductivity improves when disorder increases.

{\bf The Hamiltonian}
We consider the antiferromagnetic spin-$\frac{1}{2}$ Heisenberg model with nearest-neighbor interactions,
\begin{equation}
 \label{eq:H-spin}
 H=J \sum_{\langle i,j \rangle} {\mathbf S}_i \cdot {\mathbf S}_j,\quad J>0,
\end{equation}
where $\langle i,j\rangle $ are pairs of sites that are linked by an edge of the tiling. Since the structures we consider are constructed from four sided polygons (the thick and the thin rhombus), the model is bipartite (i.e. can be divided into two sublattices, A and B). The antiferromagnetic couplings are thus unfrustrated. At T=0, since there is no frustration in the model, one expects a N\'eel ordered ground state, with equal and oppositely directed sublattice magnetizations. This ground state expected to slightly deviate from the ground state of classical spins on the same vertices. The ground state of classical spins is $\langle S_i\rangle = \pm \frac{1}{2}$, (+) on a sublattice and (-) on the another sublattice. This deviation is expected to be small and one could calculate this deviation using linear spin wave theory (\cite{wessel, penroselong} and references therein). At finite temperature, the long range order will be destroyed due to the Mermin-Wagner-Hohenberg theorem \cite{mwh}, although short range correlations will persist. In our calculations we consider rectangular (Taylor) approximants of the Penrose tiling. These are finite samples of $N$ spins satisfying periodic boundary conditions. Approximants corresponding to the perfect deterministic Penrose structure were first obtained by a standard method and then these were randomized using the procedure described below. 

\begin{figure}[h]
\begin{center}
\includegraphics[scale=0.5]{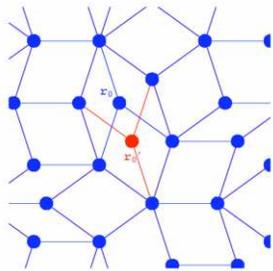}
\caption{A single phason flip showing the original ($r_0$) and final positions ($r'_0$). The original sites and bonds are shown in blue and the new site and new bonds in red.} \label{phasonflip.fig}
\end{center}
\end{figure}

{\bf Phason-flip}
A "phason flip" is a process in which a 3-fold site hops to a new allowed position. In the process, the three bonds linking it to its original nearest neighbors are effaced, while new connections appear to three new nearest neighbors (Fig.~\ref{phasonflip.fig}). Note that the original site and the new phason shifted site belong to different sublattices. To maintain a ground state of total spin zero, our random selection of phason flip centers is done with the requirement that there be an equal number of flips on each of the sublattices A and B. In our random tiling generating procedure, each step randomly selects one three-fold site from a list of possible candidates for phason flips. The new connectivity matrix is determined and the set of sites available for the next phason flip is determined. For measuring the disorder in a given value of phason flips, we calculating the overlap between the original and the phason disordered tiling. We calculate the number of shifted sites and this number normalized with the system size will be the degree of disorder ($\Delta$) for a given number of phason flips. The system sizes that we considered are N=644, 1686 and 4414 spins. The number of realizations  considered range from 100 for the smallest size, to 10 realizations in the biggest size.

\begin{figure}[h]
\begin{center}
\includegraphics[scale=0.6]{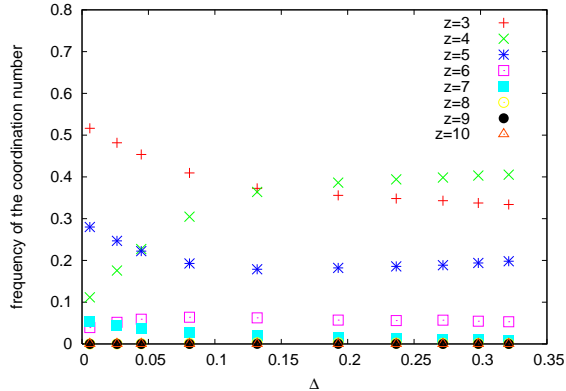}
\caption{Dependence of the coordination number distribution from the degree of disorder ($\Delta$). On the y axis: the ratio of the number of the sites with a given coordination number per the system size in the case of certain degree of disorder. (The system size we considered is N=4414.)} \label{cn.fig}
\end{center}
\end{figure}

{\bf Geometrical effects}
Phason flips clearly lead to a modification of local environments on the tiling, and in particular, the apparition of new local environments that were not present in the original tiling such as 8-fold, 9-fold and 10-fold vertices. Furthermore, finite-sized regions of perfect five-fold rotational symmetry progressively disappear with increasing degree of disorder. 
We have made a study of the evolution of the frequency of sites of each coordination number $z=3,....,10$ as a function of the degree of disorder. As shown in Fig.~\ref{cn.fig}, the disorder clearly increase the number of 4-fold sites at the expense of the 3-fold and the 5-fold sites. The original perfect structure has a large number of 3-fold and 5-fold sites, with the most common cluster found in the original perfect Penrose tiling, the "football"-shape five-fold stars. These clusters are destroyed progressively by disorder. The evolution of coordination numbers will be significant for the qualitative analysis that we will present below for the staggered magnetization.

{\bf Linear spin wave analysis}
Once the randomized samples are obtained, a linear spin wave analysis is carried out as described in \cite{wessel, penroselong}. The A and B sublattice spin operators are transformed using the Holstein-Primakoff transformation to boson operators $a_i$, $b_j$ $(i,j=1,...,N/2)$. The linearized Hamiltonian in the boson operators was then diagonalized numerically.
Once the eigenmodes and spectrum have been determined, one can find the ground state energy, and local staggered magnetizations for each realization of disorder, and carry out the statistical analyses of the results. We now describe the results for the ground state energy, the energy spectrum and local staggered magnetizations.

\begin{figure}[h]
\begin{center}
\includegraphics[scale=0.6]{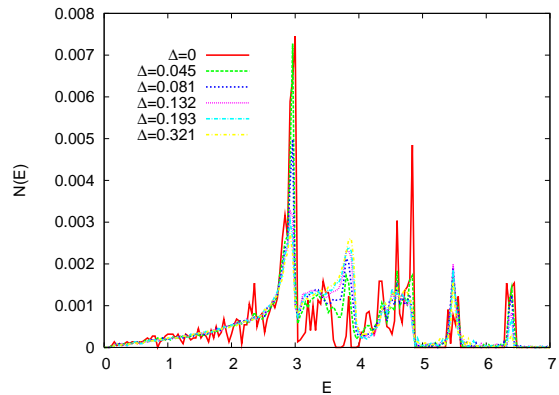}
\caption{Density of states plotted in the case of different degree of disorder: $\Delta$ (The system size is N=4414).} \label{DOS.fig}
\end{center}
\end{figure}

\begin{figure}[h]
\begin{center}
\includegraphics[scale=0.6]{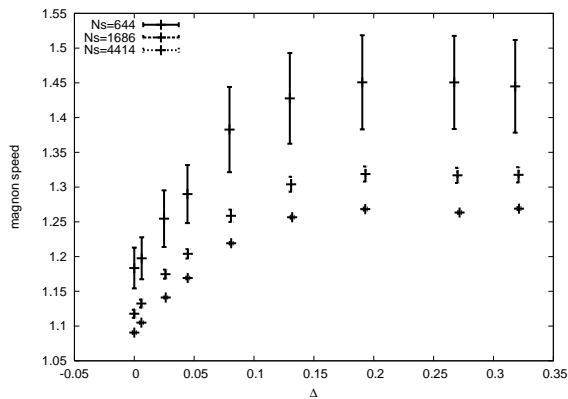}
\caption{Spin wave velocity as a function of disorder ($\Delta$). The system sizes we considered are 644, 1686 and 4414.} \label{magnon_speed.fig}
\end{center}
\end{figure}

{\bf Energy spectrum}
We show the density of states (DOS) curves as a function of the energy is expressed in the unit of J for systems with different degree of disorders (Fig.~\ref{DOS.fig}). The primary effect of increased phason disorder is to smoothen the DOS curve and fill in the gaps. The low energy tail of the integrated curve is quadratic, and can be fitted to give a spin wave velocity that increases with the degree of disorder (Fig.~\ref{magnon_speed.fig}). This indicates that spin wave propagation is facilitated by the phason disorder, in analogy with the problem of quantum diffusion of electrons in the tight binding model in quasiperiodic tilings \cite{sire}. Changes in the DOS curve  are quite small below the degenerate peak $E=3$ -- disorder only weakly affect the ``long wavelength" wavefunctions. The degenerate peak at $E=3$ is reduced due to the progressive disappearance of 3-fold sites, as mentioned already. The DOS increases with degree of disorder between energies 3.4 and 4.3, which is a range of energy corresponding to wavefunctions located mostly on four-fold sites. 

\begin{figure}[h]
\begin{center}
\includegraphics[scale=0.6]{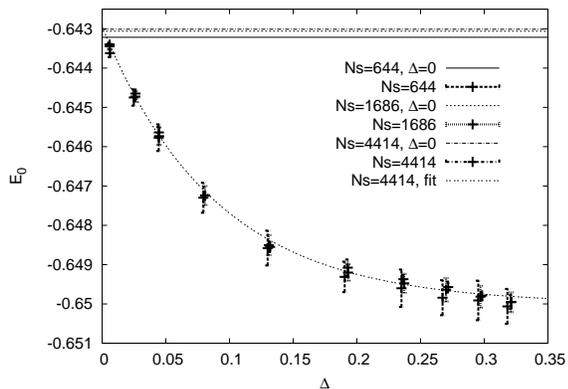}
\caption{Ground state energy as a function of increasing disorder ($\Delta$). The straight line is the ground state energy for the perfect Penrose tiling.} \label{GSE.fig}
\end{center}
\end{figure}

{\bf Ground state energy} 
Fig.~\ref{GSE.fig} shows the decrease of the ground state energy per site $E_0$ as a function of increasing degree of disorder. The decrease of $E_0$ is similar to the decrease in the average value of the staggered magnetization (see below), showing that the quantum fluctuations are increased in the random tiling as compared to the perfect case.

\begin{figure}[h]
\begin{center}
\includegraphics[scale=0.8]{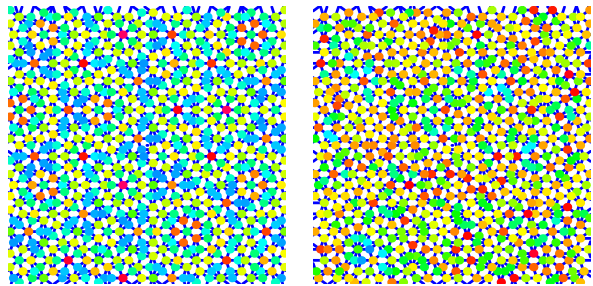}
\caption{Comparison of the real space local staggered magnetization values on the Penrose tiling approximant. We colorized the sites on the Penrose tiling using different colors for different local staggered magnetization value. The color code is the following: red, orange, yellow, green, blue with local staggered magnetization from the lowest to the highest values. The system size is 1686.
} \label{comparemags.fig}
\end{center}
\end{figure}

\begin{figure}[h]
\begin{center}
\includegraphics[scale=0.6]{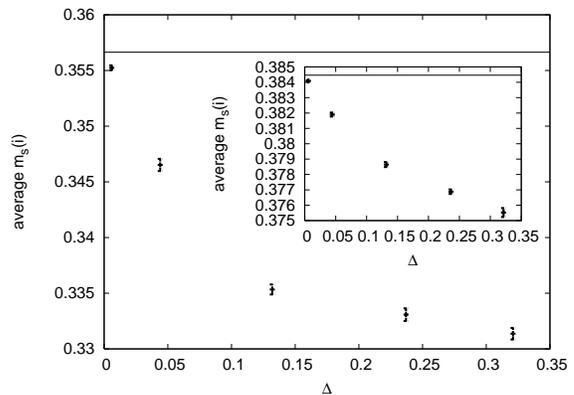}
\caption{Average local staggered magnetization in the case of different degree of disorder ($\Delta$). The straight line is the average local staggered magnetization for the perfect Penrose tiling approximant. We considered the expectation values for the thermodynamic limit. In the inset: local staggered magnetization on the central site of a two level Heisenberg-star cluster. We use the same coordination number distribution like the whole tiling has in the case of different degree of disorder ($\Delta$). The straight line is the average local staggered magnetization for the two level Heisenberg-star cluster using the same coordination number distribution like the perfect Penrose tiling.} \label{mws_comp_nr_inf.fig}
\end{center}
\end{figure}

{\bf Local staggered magnetizations}
The real space staggered magnetizations for a randomized system  $m_{si}=\vert\langle S_i\rangle\vert $ are shown in a color plot on the right hand side figure in (Fig.~\ref{comparemags.fig}).  The right hand figure shows the corresponding quantities on the perfect tiling, for comparison. The perfect case shows symmetries that the random case does not. In addition, the overall (site-averaged) staggered magnetization is clearly smaller for the disordered case (the color coding shows a redder tone, indicating smaller values). Fig.~\ref{mws_comp_nr_inf.fig} shows the average staggered magnetization as a function of degree of disorder. This quantity decreases with increasing disorder, indicating greater quantum fluctuations.

The results for the staggered magnetizations can be qualitatively explained in terms of a finite cluster model \cite{penro}. In this approximation, the local environment is retained up to second neighbor level in the spin wave calculation. Boundary conditions, and connectivity beyond second neighbors are thus neglected. Nevertheless, with this simple approximation many of the results can be obtained, using data about nearest neighbor and next nearest neighbor coordination number as input structural parameters.  We conclude from this that the local staggered magnetization values are dependent mainly on the local environments for the perfect as well as the disordered Penrose tilings (Fig.~\ref{mws_comp_nr_inf.fig} inset).

{\bf Conclusion}
We investigated the effects of randomly located "phason" flips on a Heisenberg antiferromagnet in approximants of the Penrose tiling.  We calculated the energy spectrum, the ground state energy and local staggered magnetizations in the ground state using linear spin wave approximation in order to investigate the effects of increased phason disorder. The magnon speed estimated from the low energy part of the spectrum increases with phason disorder. The ground state energy is observed to decrease with increasing disorder. We investigated the changes of the staggered magnetizations in real space. The ground state loses its symmetry properties and the overall staggered magnetization decrease with increasing degree of disorder. This observation, along with the decrease of the ground state energy, show that disorder tends to increase quantum fluctuations in this system.

\end{document}